# Effect of Synchrotron Polarization in Grazing Incidence X-ray Fluorescence analysis

Gangadhar Das, M K Tiwari*, A K Singh, and Haranath Ghosh
*Indus Synchrotrons Utilisation Division,
Raja Ramanna Centre for Advanced Technology, Indore-452013, India.*
*Corresponding author, email: mktiwari@rrcat.gov.in

**Abstract**: Total reflection x-ray fluorescence (TXRF) spectroscopy has seen a remarkable progress over the past years. Numerous applications in basic and applied sciences prove its importance. The large spectral background which is a major detrimental factor in the conventional x-ray fluorescence technique, limits the element detection sensitivities of the technique to µg/g (ppm) range. This spectral background reduces to a great extent in the TXRF technique due to the low extinction depth of the primary incident x-ray beam. In synchrotron radiation (SR) based TXRF measurements the spectral background reduces further because of the polarization of the synchrotron x-ray beam. Here, we discuss in detail the influence of synchrotron polarization on the spectral background in a fluorescence spectrum and its significance towards TXRF detection sensitivities. We provide a detailed theoretical description and show that how anisotropic scattering probability densities of the Compton and Elastic scattered x-rays depend on the scattering angle ($\theta$) and azimuthal angle ($\phi$) in the polarization plane of the SR beam.

## Introduction

X-ray fluorescence (XRF) spectrometry is a well established and widely used non-destructive technique for the trace element analysis. The unique features of XRF technique make it a potential candidate for majority of applications in various fields of sciences. Apart from the research applications, the technique has also seen remarkable success in the industry especially in maintaining the quality control of ultra pure grade chemicals, reagents and products[1,2]. In conventional XRF technique, the element detection sensitivities are largely limited to the µg/g (ppm) range due to large spectral background produced by the Compton scattered x-rays from the specimen. TXRF is another variant of the energy dispersive-XRF (EDXRF) technique, where the complexity of large spectral background eliminates to a great extent owing to the high reflectivity on the flat surface and low penetration depth of the primary x-ray beam in the substrate material, on which the incident x-rays are allowed to impinge at glancing incidence angles. All these features improves the detection sensitivies of TXRF ~ 2-3 orders of magnitude or better than the conventional XRF, typically in the range of parts per billion (ppb) levels for most of the elements[3, 4]. Another approach to reduce the spectral background in x-ray fluorescence measurements is to utilize linearly polarized primary radiation for excitation[1]. Based on the classical dipole oscillator, it is well understood that the emission profile of scattered radiation is anisotropic in case of a polarized x-ray beam. Maximal signal to noise ratio can be realized if fluorescence signal is measured in a position, where the contribution of anisotropic scattered radiation is minimal[5].

In literature, several attempts have been made earlier that report significant reduction of the spectral background and hence improved detection sensitivities of the XRF technique. Embong et al.[6] have reported sub-ppm level detection sensitivities using three-dimensional polarization excitation geometry for multi element sediment samples by means of variation of various experimental parameters like excitation geometry, x-ray energy, secondary targets, operating voltage and current of the x-ray tube source, and thickness of filters installed in the spectrometer. Spolnik et al.[7] highlighted that the use of three-dimensional polarizing optical geometry in combination with secondary targets is highly beneficial to significantly reduce the scattered background and they achieved nearly one order of magnitude improved detection sensitivities as compared to the conventional EDXRF measurements.

In the present work, we describe the effect of the linear polarization of the SR radiation in grazing incidence x-ray fluorescence analysis. We provide a detailed mathematical description and show that how the Compton and elastic scattering probability densities depend on the scattering angle ($\theta$) and the azimuthal angle ($\phi$) in plane of synchrotron polarization. The analytical treatment presented here takes into account the effect of incident x-ray energy as well as the spatial distribution of the scattered x-ray photons. Our results show that the Compton and elastic scattered x-rays have an optimal intensity contrasts for some particular positions of $\theta$ and $\phi$ because of the anisotropic emission characteristics of the scattered radiation. We correlate our theoretical findings with the experimental results using TXRF measurements of standard reference materials. Our results reveal that one obtains approximately one order of magnitude enhancement in the spectral signal strength, if the spectroscopy detector is placed in the plane of polarization of synchrotron beam ($\theta = \pi/2^o$ and $\phi = 0^o$) in contrast to the conventional mounting geometry ($\theta = \pi/2^o$, $\phi = \pi/2^o$).

**Theoretical background**

The number density of scattered photons in a well defined solid angle strongly depends on the plane of polarization of the incident x-ray beam. However, if the exciting radiation is unpolarized then the isotropic nature of the scattered radiation allows us to choose detector collimator geometry randomly.

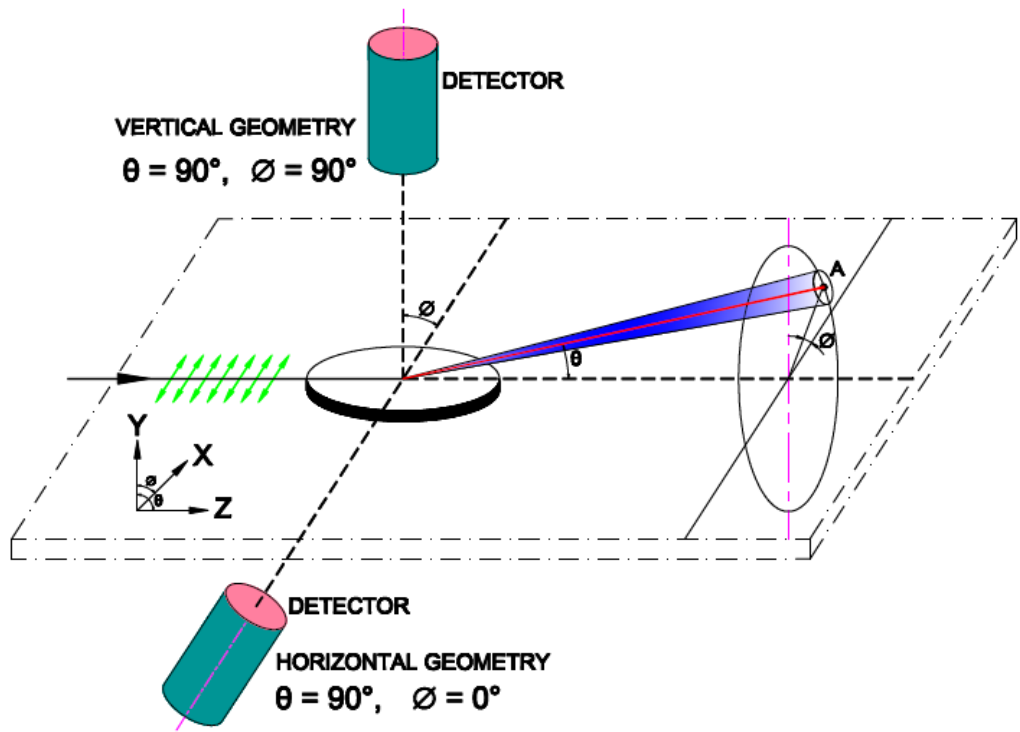

*Fig. 1 Experimental set up used in our study for the angle dependent XRF measurements. The spectroscopy detector is placed in the vertical geometry (i.e. $\theta=\pi/2^o$, $\phi=\pi/2^o$) as well as horizontal geometry (i.e. $\theta=\pi/2^o$, $\phi=0^o$) to collect the XRF spectra.*

Fig. 1 illustrates a typical geometrical configuration used in the x-ray fluorescence measurements. For our convenience we have chosen x-ray beam propagation along the Z-axis and the electric field vibrations are considered to be constrained in parallel to the X-axis (i.e. s-polarized synchrotron beam). The scattering angle (θ) and azimuthal angle (ϕ) are measured from the Z-axis and from the plane of electric filed vibrations.

The differential Compton scattering cross section can be obtained from the Klein Nishina formula[8]. In case of the linearly polarized x-ray beam of energy $E$ the differential Compton and elastic scattering cross sections as function of polar angle $\theta$ and azimuthal angle $\phi$, from a scattering centre (atomic number Z) can be expressed as

$$\frac{d}{d\Omega}\sigma_{Com}(\theta,\phi,E) = \frac{d}{d\Omega}\sigma_{KleNis}(\theta,\phi,E)S(\text{x, Z})$$

$$= \frac{r_e^2}{2}\gamma^2(\gamma + \frac{1}{\gamma} - 2sin^2\theta\, cos^2\phi)S(\text{x, Z}) \quad (1)$$

$$\frac{d}{d\Omega}\sigma_{elas}(\theta,\phi,E) = \frac{d}{d\Omega}\sigma_{Thom}(\theta,\phi,E)F^2(\text{x, Z})$$

$$= r_e^2(1 - sin^2\theta\, cos^2\phi)F^2(\text{x, Z}) \quad (2)$$

where $\frac{d}{d\Omega}\sigma_{KN}(\theta,\phi,E)$ is the Klein–Nishina differential cross section, $S(x, Z)$ is the incoherent scattering factor, $F(x, Z)$ is the atomic form factor[9], x can be defined as $Sin(\theta/2)(E/1.239852$ keV nm$)$, $r_e = e^2/4\pi\epsilon_0 m_e c^2$ represents classical electron radius, $e$ is the charge of an electron, $m_e c^2$ is the rest mass energy for an electron, $\epsilon_0$ is the permittivity of free space, and the dimensionless quantity $\gamma$ is defined as

$$\gamma = (1 + \Delta\lambda_{Compton}/\lambda)^{-1} \quad (4)$$

where $\Delta\lambda_{Compton}$ describes the wavelength shift in the Compton scattering process and $\lambda$ is the wavelength of the incident x-ray beam. The differential cross section in the classical Thomson scattering from single electron is given by

$$\frac{d}{d\Omega}\sigma_{Thom}(\theta) = (1 + cos^2\theta)\frac{r_e^2}{2} \quad \text{(Un-polarised light)}$$

$$\frac{d}{d\Omega}\sigma_{Thom}(\theta,\phi) = (1 - Sin^2\theta Cos^2\phi)r_e^2 \quad \text{(Polarised light)} \tag{3}$$

In the case of un-polarised incident radiation the scattering cross sections are given by

$$\left(\frac{d\sigma_{Comp}(\theta,E)}{d\Omega}\right)_{unpol} = \left(\frac{d\sigma_{Klenis}(\theta,E)}{d\Omega}\right)_{unpol} S(x, Z)$$

$$= \frac{r_e^2}{2}\gamma^2(\gamma + 1/\gamma - sin^2\theta)S(x, Z) \tag{5}$$

$$\left(\frac{d\sigma_{elas}(\theta,E)}{d\Omega}\right)_{unpol} = \left(\frac{d\sigma_{Thom}(\theta,E)}{d\Omega}\right)_{unpol} F^2(x, Z)$$

$$= \frac{r_e^2}{2}(1 + cos^2\theta)F^2(x, Z) \tag{6}$$

From the above equations one can find the total Compton and elastic scattering cross section for an element[10]. The geometrical distribution of the scattered photons can be obtained by calculating the scattering probability density function described by

$$f(\theta, \phi, E) = (differential\ cross\ section)/(Total\ cross\ section \tag{7}$$

The scattering probability density for a photon of energy $E$, which scatters within the finite polar angle domain of $\phi$ and $\phi + d\phi$, at a specific scattering angle $\theta$ is given as

$$f(\theta,\phi,E) = \frac{1}{\sigma(\theta,\phi,E)}\frac{d\sigma(\theta,\phi,E)}{d\Omega} = \frac{d\sigma(\theta,\phi,E)}{d\Omega}\left[\int_0^{2\pi}\frac{d\sigma(\theta,\phi,E)}{d\Omega}d\phi\right]^{-1} \tag{8}$$

To estimate the Compton and elastic scattering density profile we need to substitute the differential scattering cross sections in the eqn (8). This allows us to arrive a relation

$$f(\theta,\phi,E) = \frac{1}{2\pi}\left[1 - \frac{sin^2\theta}{G(\theta)}\cos 2\phi\right] \tag{9}$$

$$\text{where } G(\theta) = \gamma + \frac{1}{\gamma} - sin^2\theta \quad \text{(Compton scattering)}$$

$$2 - sin^2\theta \quad \text{(elastic scattering)}$$

We carried out numerical simulations to investigate the effect of angular dependence of the scattered photons. Fig.2 depicts the plot of scattering probability densities (eqn 9) for the (a) Compton and (b) elastic scattered x-rays as a function of scattering angles $\theta$ and azimuthal angle $\phi$ assuming incident x-ray energy of 15 keV.

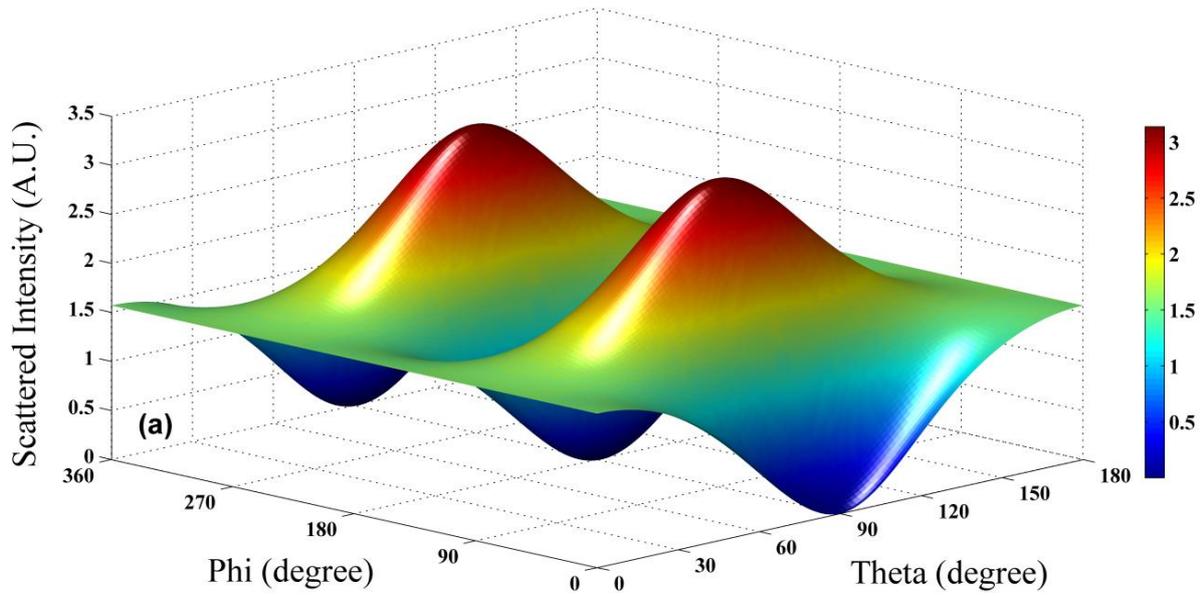

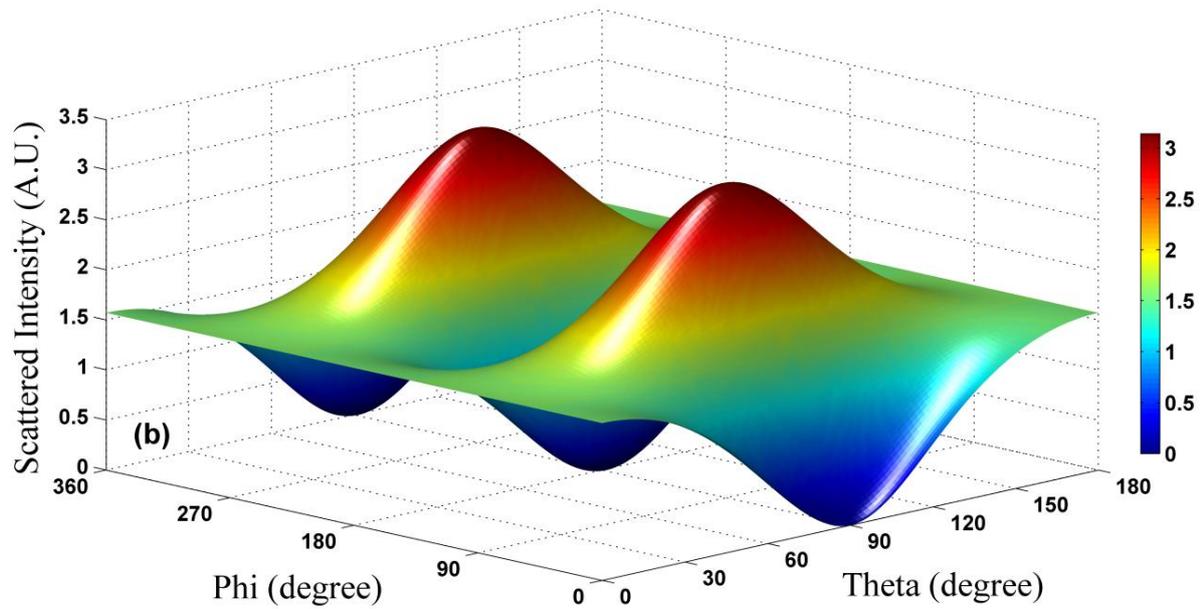

Fig. 2 *Angular distribution of the scattered photon for (a) Compton and (b) elastic scattering as a function of scattering angle (θ) and azimuthal angle (ϕ) at incident x-ray energy of 15 keV.*

It can be seen from Fig. 2a and 2b that both the Compton and elastic scattered x-ray photons show strong angular anisotropy (with respect to angles $\theta$ and $\phi$). The scattered intensity has its maxima at angles $\theta=\pi/2^o$, $\phi=\pi/2^o$ and minima at $\theta=\pi/2^o$, $\phi=0^o$ herein referred as vertical geometry and horizontal geometry respectively. This implies that if the spectroscopy detector is placed in the horizontal geometry (i.e. $\theta=\pi/2^o$, $\phi=0^o$) one will observe minimum spectral background in the fluorescence spectrum.

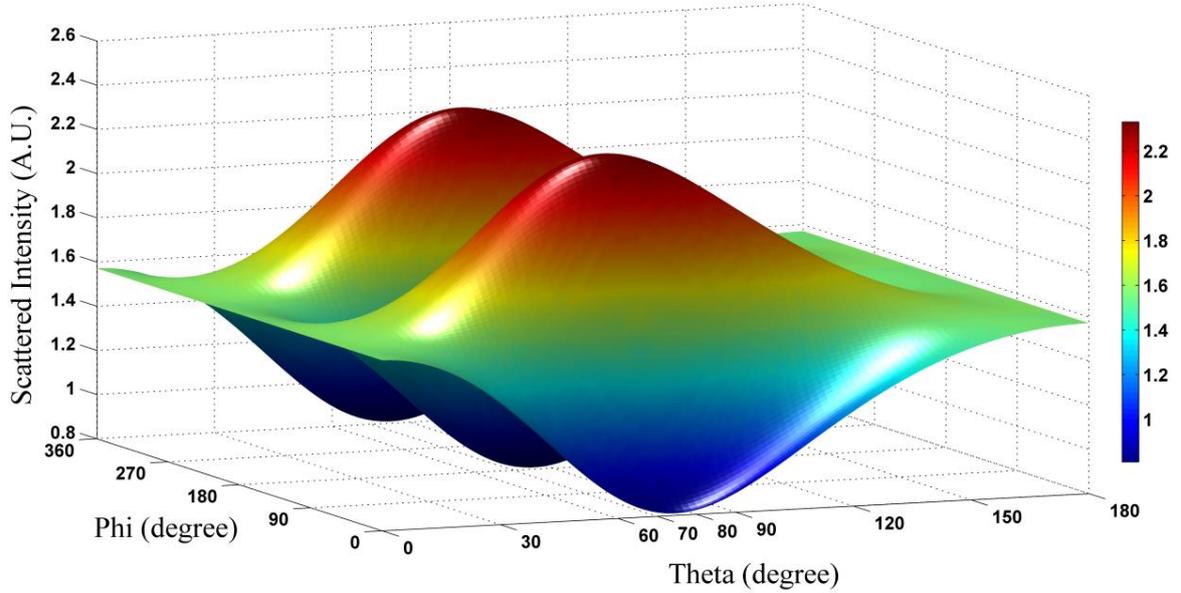

Fig. 3 *Angular distribution of the Compton scattered photon as a function of scattering angle (θ) and azimuthal angle (ϕ) at the 1MeV monochromatic incident x-ray beam.*

We have also evaluated the behavior of angular dependence of the scattered photons at higher incident x-ray energies. It has been realized from simulations that energy dependence does not have much influence on the angular distribution of scattered photons especially in the medium energy range (10-100 keV). However, if the energy of the incident x-ray photon is greater than 100 keV then angular profile of the scattered photon changes considerably. Fig. 3 shows computed angular distribution profile for the Compton scattered x-rays as a function of $\theta$ and $\phi$ at incident x-ray energy of 1 MeV. It can be observed from this figure that minima and maxima of the Compton scattering distribution are shifted towards the lower scattering angle side ($\theta \approx 75°$) but with same value of the azimuthal angles ($\phi$).

Conversely the angular distribution of elastically scattered intensity more or less remains unchanged at lower as well as at higher incident x-ray energies.

The scattering measurements also allows us to determine degree of polarization $P$ using the following formula Perrin

$$P = \frac{I_\perp - I_\parallel}{I_\perp + I_\parallel} \tag{10}$$

where $I_\perp$ and $I_\parallel$ the intensity of the scattered radiation (Compton or elastic) in the parallel and perpendicular planes with respect to incident electric field[11].

**Experimental**

The total reflection x-ray fluorescence measurements on standard samples were carried out at the microfocus X-ray fluorescence (BL-16) beamline of the Indus-2 synchrotron radiation facility. The BL-16 beamline is designed to work in the x-ray energy range of 4 – 25 keV. The details of the beamline and different experimental stations installed on this beamline are described elsewhere[12]. TXRF station of the BL-16 beamline comprises of an in-house developed 4-axes motorized sample manipulator installed on a vibration free optics table. It provides vertical as well as horizontal movement to a sample with respect to the x-ray beam with a minimum step size resolution of $\approx$ 1 micron. Two independent angular motions (Tilt and Roll) are also provided for the alignment of the sample as well as to set small grazing incidence to achieve total external reflection condition on sample reflector surface with a minimum step size resolution of ~ $0.005^0$. Monochromatic x-rays of energy 15keV monochromatized from a Si (111) double-crystal monochromator was used for excitation of the samples at grazing incidence angles. Provisions have been made to mount the solid state spectroscopy detector (*Vortex, USA*) in horizontal as well as vertical plane to the substrate surface (see Fig. 1). The spectroscopy detector has an active surface area of 50 mm$^2$ and an energy resolution of $\approx$ 140 eV at 5.9keV (Mn K$\alpha$) x-rays. An x-ray CCD camera was also employed to confirm the TXRF condition by observing reflected x-ray beam from the sample substrate. For the TXRF investigations we used NIST-1640 standard reference material (*trace elements in natural water)* and an ICP-IV Merck multielement standard. Aliquots of ~ 20 µL volume of standard were micropipette on polished Si (100) substrates. Instead of using a single element standard we used multielement reference materials to investigate sensitivities of the different elements over the full spectral background.

**Results and discussions**

Fig. 4 comprise measured TXRF spectra of the NIST-1640 reference material at 15 keV monochromatic x-ray energy in the vertical ($\theta=\pi/2^o$, $\phi=\pi/2^o$) and horizontal ($\theta=\pi/2^o$, $\phi=0^o$) geometries. The figure clearly confirms the presence of different elements in the NIST reference material of atomic number Z ranging from 19 (K) to 33 (As). The peaks of Ar and Si are observed from the Si substrate and atmosphere respectively.

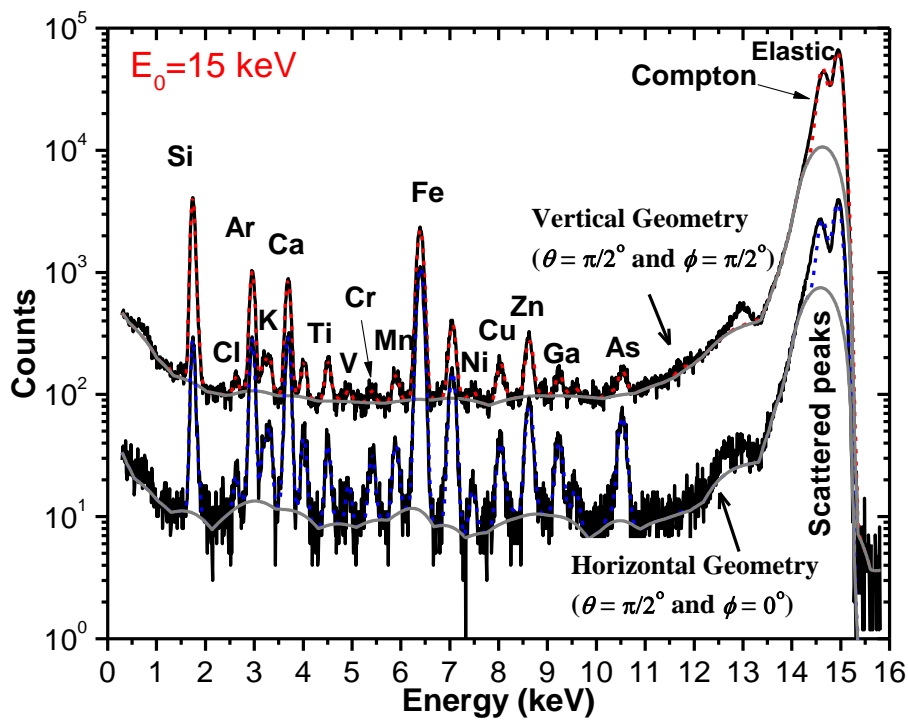

Fig. 4 *Measured TXRF spectra of the NIST-1640 reference material in vertical geometry ($\theta=\pi/2^o$, $\phi=\pi/2^o$) and horizontal geometry ($\theta=\pi/2^o$, $\phi=0^o$). Here, Ga (100 ppb) used as an internal standard reference element. Black solid lines are experimental data, dotted red and blue lines are the fitted data*

*in the two measurement geometries. The thick grey lines represent a good estimation of the spectral background.*

It can be noticed from the Fig. 4 that signal to background ratio is unambiguously higher in the horizontal geometry compared to the vertical geometry of the detection setup. This in turn allows us detection of an analyte with an encouraging level of confidence limit which is present at very low concentration in the NIST specimen (for example As- 26.67 ppb). In case of vertical detection geometry, the peak height for the same element is very small compared to the horizontal geometry. This increased signal potency mainly attributes to the reduction of scattered background (Compton and elastic) originated from the specimen. To validate our observations we have performed similar measurements using another reference standard material (ICP-IV sample) containing different elements at ppm level.

Fig. 5(a) reports measured TXRF spectrum of an ICP-IV ) standard reference material containing 50.53 ppm of each elements at 15 keV monochromatic x-ray energy in the vertical geometry ($\theta=\pi/2^o$, $\phi=\pi/2^o$) as well as horizontal geometry ($\theta=\pi/2^o$, $\phi=0^o$). From this figure it can be seen that the effect of spectral background is not very significant as recognized in the NIST standard reference sample because most of the elements in the ICP-IV standard sample are present in the ppm range. Nonetheless we could observe the same kind of behaviour of the spectral background for the ICP-IV standard too. The effect spectral background on the fluorescence signal becomes obvious if we give attention to TI L-$\gamma_1$ peak. The fluorescence peak of TI L-$\gamma_1$ could only be observed clearly, as a result of significant decrement of the Compton scattered background in the horizontal geometry as shown in Fig. 5(b). This suggests that the influence of the spectral background on the fluorescence signal is decisive if the analytes (in ppb level) are present at the tailing background of the Compton scattered peak. These results are consistent with the observations of Kenji Sakurai et al.[13].

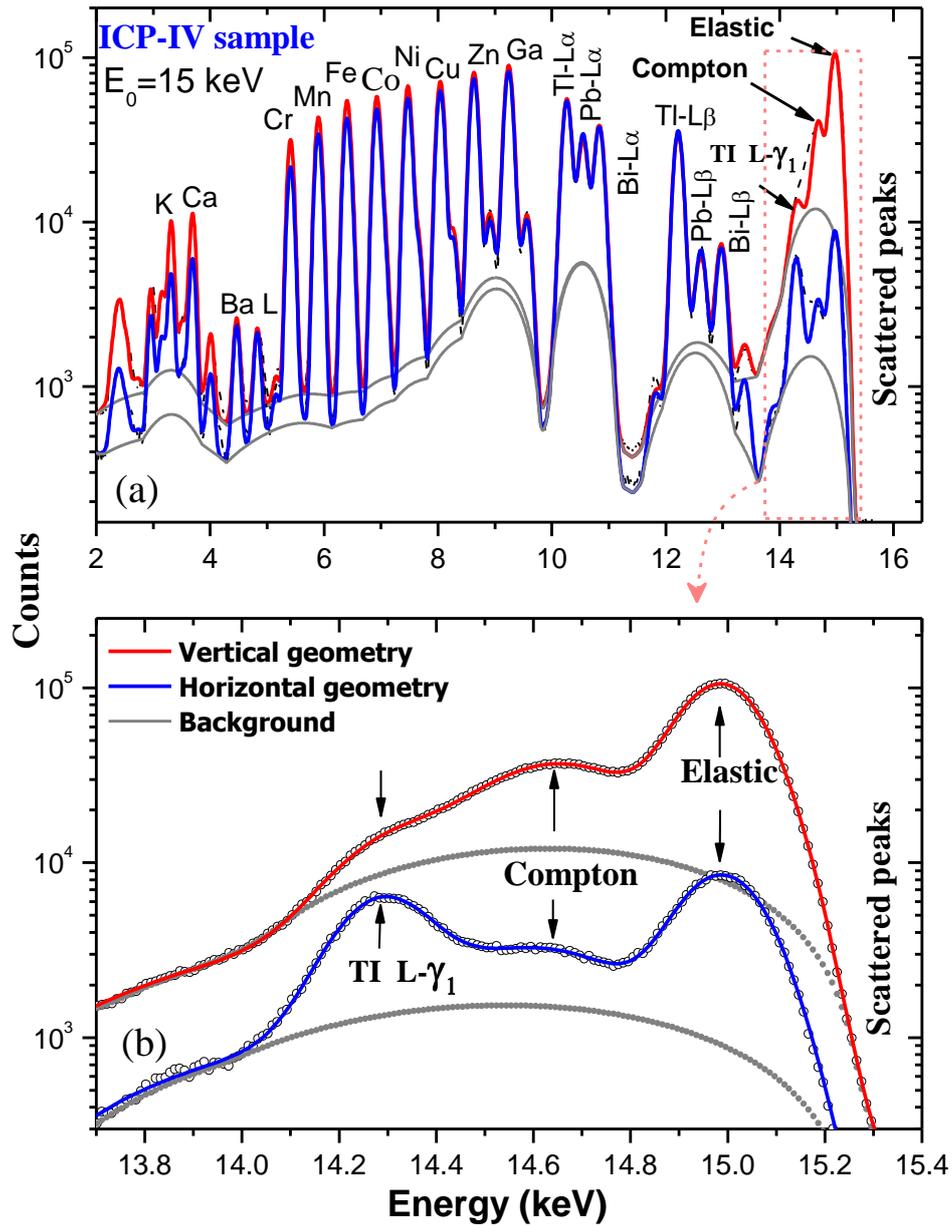

Fig. 5 (a) *Recorded TXRF spectra for an ICP-IV Merck reference material containing 50.35 ppm wt. concentration of each elements in vertical geometry ($\theta=\pi/2^o$, $\phi=\pi/2^o$) and horizontal geometry ($\theta=\pi/2^o$, $\phi=0^o$). In this figure black dotted lines are experimental data, solid red and blue lines are fitted data in the two geometries, and the thick grey lines represent a good estimation of the spectral background.* (b) *Extended view of the encircle region of Fig. (a), where Tl-L$\gamma_1$ peak is individually fitted in the presence of large Compton scattered background. It can be seen that visibility of the Tl-L$\gamma_1$ peak gets enhanced in the horizontal detection geometry due to the reduced spectral background.*

To speak out more quantitatively we calculate relative signal strength in two measurement geometries for the NIST sample. Here we define the relative signal strength as a ratio of signal to noise ratios (S/N = net area intensity/background) in horizontal to the vertical geometries.

*Table-1 Geometrical dependence of Relative signal strength for the NIST material.*

| Element | Z (Atomic Number) | S/N (Vertical geometry) | S/N (Horizontal geometry) | Relative signal strength |
|---|---|---|---|---|
| K | 19 | 0.43 | 1.55 | **3.6** |
| Ca | 20 | 2.99 | 9.80 | **3.3** |
| V | 23 | 0.02 | 0.10 | **7** |
| Cr | 24 | 0.13 | 1.15 | **9.4** |
| Mn | 25 | 0.30 | 1.38 | **4.7** |
| Fe | 26 | 7.33 | 36.00 | **5** |
| Ni | 28 | 0.13 | 0.90 | **6.7** |
| Cu | 29 | 0.45 | 1.83 | **4** |
| Zn | 30 | 0.72 | 3.12 | **4.4** |
| Ga | 31 | 0.25 | 1.60 | **6.6** |
| As | 33 | 0.30 | 2.29 | **7.5** |

In the table -1we have tabulated the relative signal strength for different elements. The tabulated data clearly show that the signal strength enhancement in horizontal geometry is almost one order of magnitude higher compared to the vertical detection geometry. This increased signal strength greatly influences the detection sensitivities of various elements in TXRF excitation. The minimum detection limits (MDLS) obtained for different elements using the formalism

$$C_{DL} = \frac{3\sqrt{I_B}}{I_A/C_A} \qquad (11)$$

where $I_B$ is the integrated linear background intensity under the fluorescence peak of analyte element A, $I_A$ is the net area intensity of the analyte and $C_A$ is the weight concentration of analyte.

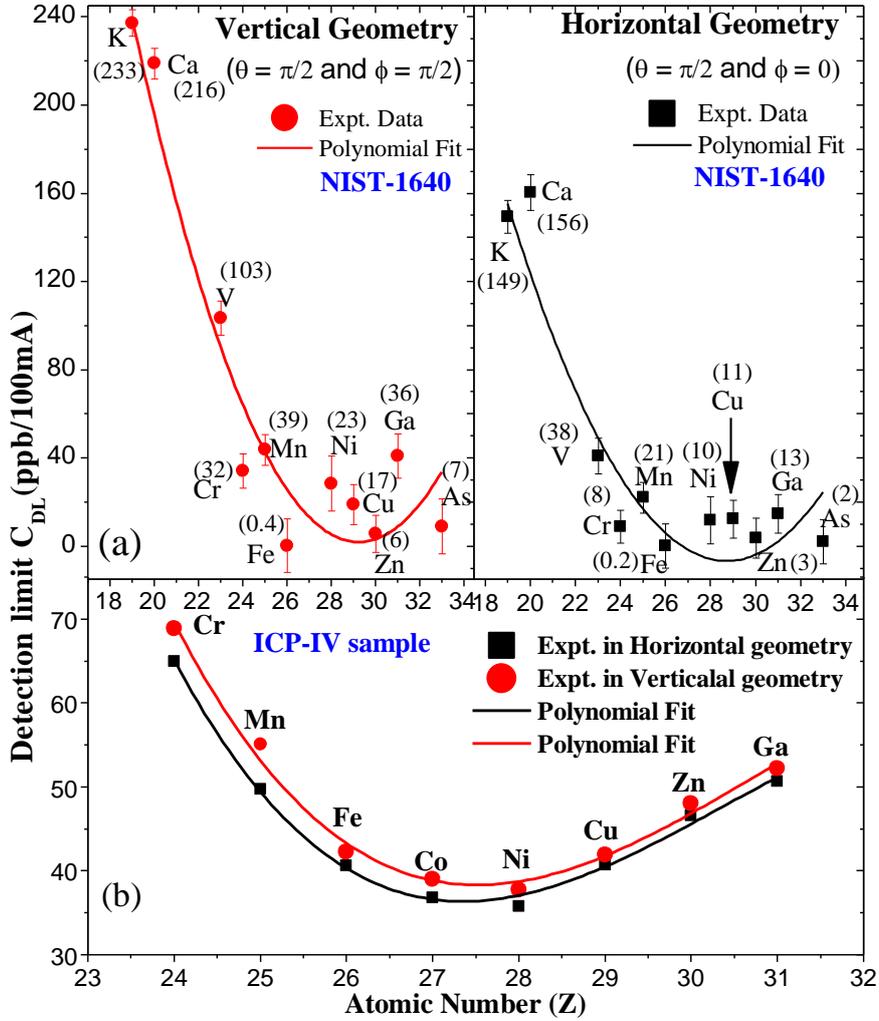

Fig. 6 *Variation of elemental detection sensitivities for (a) NIST-1640 (in ppb) and (b) ICP-IV (in ppm) standard reference samples as function atomic number Z in the vertical ($\theta=\pi/2^o$, $\phi=\pi/2^o$) and horizontal ($\theta=\pi/2^o$, $\phi=0^o$) detector mounting geometry.*

Fig.6 shows the measured detection sensitivities of various elements in the vertical and horizontal geometries for the (a) NIST-1640 and (b) ICP-IV standard reference samples. From the Fig.6 (a) it can be seen that detection sensitivities in the horizontal geometry range in ~ 234 ppb to 2 ppb for elements of atomic number Z ranging from 19 to 33 for the NIST-1640 standard reference material. On the other hand, in the vertical detection geometry these detection sensitivities have been found to be deteriorated by a factor of 2. Furthermore, we obtained an analogous conclusion in the case of ICP-IV standard reference sample. The detection sensitivities of various elements in the horizontal detection geometry were found slightly superior compared to the vertical detection geometry as depicted in Fig. 6(b). The higher $C_{DL}$ values were found for different elements in case of ICP-IV sample as compare to the NIST-1640 sample because of higher concentration of analyte present in the ICP-IV sample. The minimum detection sensitivity of an element in the x-ray fluorescence technique strongly depends on the analyte mass present in the specimen[14].

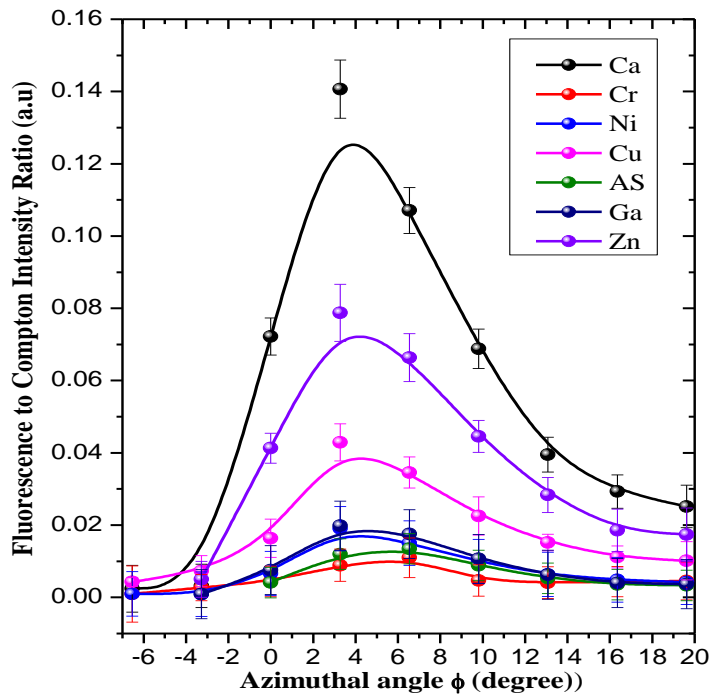

*Fig. 7 Variation of fluorescence signal to scattered Compton intensity ratios for different elements in the NIST-1640 standard reference material as a function of azimuthal angle (ϕ).*

These findings are consistent with the experimental observations reported by Margui[15]. They showed that the quantification of *Cadmium* in 100mL aqueous samples in the range of 1-20 μg is only practicable by using polarized energy dispersive x-ray fluorescence instrument (P-EDXRF) in contrast to the conventional EDXRF and WDXRF spectrometer.

For precise determination of the detector geometry in the horizontal plane we performed further TXRF measurements at different shallow azimuthal angles. Fig.7 shows variation of fluorescence to Compton intensity ratios for different elements in the NIST 1640 standard reference material for azimuthal angles ranging from $-7^o$ to $20^o$. It may be noted from the Fig.7 that the maximum of the fluorescence to Compton intensity ratio lies in the angular region of $\phi= 3^o$ $to$ $\phi=6^o$. This azimuthal angular span is basically delectated by the active area of the detector element and detector-to-sample distance. In our case we placed the solid state spectroscopy detector ~ 30mm apart from the sample. We obtain maximum signal strength (signal to noise ratio) for the detector take off angle of $\phi \sim 5^0$.

We have also estimated the degree of polarization using the Compton and coherently scattered radiations in the TXRF spectrum. The gross areas of the Compton and elastic scattered peaks were measured in the vertical and horizontal detection geometries. The polarization factor then calculated following the eqn (10) and values are tabulated in table- 2. We have observed degree of polarization ~ $88\pm 2$ % for the synchrotron x-ray beam for the TXRF excitation at the Bl-16 beamline.

*Table-2 Measured degree of Polarization for SR beam at the BL-16 beamline.*

| Radiation | $I_\perp$ | $I_\parallel$ | Polarization($P$) |
|---|---|---|---|
| Compton | 3182085 | 212829 | 0.87 |
| Elastic | 1563135 | 95212 | 0.88 |

**Conclusion**

We have analyzed the effect of synchrotron beam polarization in grazing incidence x-ray fluorescence measurements. We have shown that the scattering probability densities of the Compton and elastic scattered x-rays strongly depend on the scattering angle ($\theta$) and azimuthal angle ($\phi$) in the polarization plane of the synchrotron beam and in some extent it also depends on the energy of the incident radiation. The anisotropic nature of the scattered radiation imposes a constraint to place an x-ray spectroscopy detector in a condition where the contribution of the scattered background is minimal. Thus, the choice of detector collimator geometry is of critical importance and strongly influences the TXRF detection sensitives. Our theoretical results closely corroborate with the experimental TXRF measurements carried out using NIST-1640 and ICP-IV standard reference materials. Our results reveal that one obtains approximately one order of magnitude better signal to background ratio in the horizontal detection geometry in contrast to vertical detection geometry. As a result, the TXRF detection sensitivities found to be improved by ~20 – 60% for many elements in the horizontal detection geometry relative to the vertical geometry. The specificity of the angular selectivity of detector collimator geometry was also studied by varying azimuthal angle. For our TXRF setup we obtained best signal to Compton background contrast in the angular region of $\phi=2^o$ to $\phi=8^o$.

It may be important to mention that our results provide a general guideline for the mounting of spectroscopy detector especially for the synchrotron based grazing incidence x-ray fluorescence and total reflection x-ray fluorescence measurements to achieve better signal to spectral background contrast and will be of vital importance to the field ultra-trace element analysis.

**References**

1. R. E. Van Grieken and A. Markowicz, *Handbook of X-ray spectrometry,* Marcel Dekker, 1993, pp. 181.

2. M. K. Tiwari, G. S. Lodha, K. J. S. Sawhney, B. Gowrisankar, A. K. Singh, G. M. Bhalerao, A. K. Sinha, Arijeet Das, A. Verma and R. V. Nandedkar, C*urrent Science*, 2008, **95**, 603-609.

3. Carlos A. Perez, Martin Radthe, Hector J. Sanchez, X-ray Spectrum, 1999, **28,** 320-326.

4. M K Tiwari, A K Singh, Gangadhar Das, Anupam Chowdhury, and G S Lodha, Synchrotron Total Reflection X-ray Fluorescence at BL-16 Microfocus Beamline of Indus-2, AIP Conference Proceedings, 2014, **1591**, 642-643.

5. F. Meirer, A. Singh, G. Pepponi, C. Streli, T. Homma, P. Pianetta, *J. Anal. At. Spectrom.,* 2010, **29**, 479-496; C. Streli, P. Wobrauschek, F. Meirer and G. Pepponi, *J. Anal. At. Spectrom.,* 2008, **23**, 792–798.

6. http://www.iaea.org/inis/collection/NCLCollectionStore/_Public/29/041/29041545.pdf

7. Z. Spolnik , K. Belikov, K. Van Meel, E. Adriaenssens, F. De Roeck, R. Van Grieken, Applied Spectroscopy, 2005, **59,** 1465-1469.

8. Klein and Y Z Nishina, Z. Phys., 1929, **52,** 853.

9. Douglas E Peplowy and Kuruvilla Verghese, Phys. Med. Biol., 1998, **43,** 2431–2452.

10. P Latha, K K Abdullah, M P Unnikrishnan, K M Varier, and B R S Babu, Physica Scripta, 2012, **85**, 035303.

11. Perrin Francis, J Phys Radium., 1926, **7**, 390–401.

12. Tiwari M K, Gupta P, Sinha AK, Kane SR, Singh AK, Garg SR, Garg CK, Lodha GS, Deb SK, *J. Synchrotron Rad.,* 2013, **20**, 386–389.

13. Kenji Sakurai , Hiromi Eba ,and Shunji Goto , Japanese Journal of Applied Physics, 1998, **38**, 332-335.

14. Tiwari M K, A K Singh, and K J S Sawhney, Analytical Sciences, 2005, 21, 143-147.

15. E. Margui, B. Zawisza, R. Skorek, T. Theato, I. Queralt , M. Hidalgo, R. Sitko, Spectrochimica Acta Part B, 2013, **88**, 192-197.